\documentstyle[12pt,epsf]{article}
\newtheorem{thm}{Theorem}

\newtheorem{lem}{Lemma}
\newtheorem{df}{Definition}

 \newcommand{\Integer}{\:\mbox{\sf Z} \hspace{-0.82em} \mbox{\sf Z}\,}
  
 \newcommand{\Real}{\mbox{I \hspace{-0.82em} R}}

 \newcommand{\Z}{\Integer}

 \newcommand{\GL}{gap labelling}

 \newcommand{\bd}{\begin{df}}
 \newcommand{\bt}{\begin{thm}}
 \newcommand{\bl}{\begin{lem}}
 \newcommand{\ed}{\end{df}}
 \newcommand{\be}{\begin{equation}}
 \newcommand{\ee}{\end{equation}}
 \newcommand{\bi}{\begin{itemize}}
 \newcommand{\ei}{\end{itemize}}
 \newcommand{\et}{\end{thm}}
 \newcommand{\el}{\end{lem}}
 \newcommand{\bs}{\bigskip}
 \newcommand{\ms}{\medskip}
 \newcommand{\eb}{\hfill$\Box$}
 \newcommand{\tl}{{\cal T}}

 \newcommand{\pkt}{puncture}

 \newcommand{\Ch}{\chi^{}}
 \newcommand{\A}{{\cal A}}
 \newcommand{\2}{^{pct}}
 
 \newcommand{\mi}{pattern}
 \newcommand{\mii}{oriented pattern class}
 
 \newcommand{\ti}{tile}

 \newcommand{\Om}{\Omega}

 \newcommand{\pf}{{\cal P}_{\Sigma}}
 
 \newcommand{\AF}{{\cal A}_{\Sigma}}

 \newcommand{\Gr}{{\cal R}}
 
 \newcommand{\So}{Schr\"odinger operator}

 \newcommand{\sst}{substitution}

 \newcommand{\fl}{substitute}
 
 \newcommand{\ncg}{non commutative geometry}
 \newcommand{\saum}{border}
 \newcommand{\CA}{$C^*$-algebra}

\newcommand{\E}{E_\tl}
\newcommand{\C}{C(\Om,\Z)}
\newcommand{\KT}{C(\Om,\Z)/\E}
\newcommand{\6}{1}

\newcommand{\KF}{\C/E_\Sigma}
\newcommand{\saup}[2]{F^{#1}(#2)}

\newcommand{\eg}{\alpha}

 \title{
 \bf
 On $K_0$-Groups for Substitution Tilings
   \vspace{1.5em}}
 \author{Johannes Kellendonk}
 \date{Department of Mathematics, King's College London,\\
       Strand, London WC2R 2LS\\
        \vspace{.2em}
       {\small E-mail: johannes@mth.kcl.ac.uk}}

\begin{document}
 \maketitle

 \begin{abstract}
\noindent
The group $C(\Om,\Z)/\E$ is determined for tilings which are invariant
under a locally invertible primitive \sst\ which forces its \saum.
In case the tiling may be obtained by the generalized
dual method from a regular grid this group furnishes part of the $K_0$-group
of the algebra of the tiling. Applied to Penrose tilings one obtains
$K_0(\A_\tl)=\Z^8\oplus\Z$.
 \end{abstract}
 \begin{flushright}
 \parbox{12em}
  { \begin{center}
      KCL-TH-95-1
 \end{center} }
 \end{flushright}

\newpage

\bibliographystyle{unsrt}

\section*{Introduction}

Non commutative geometry furnishes a framework for the discussion of tilings.
Part of the topological aspect of \ncg\ is the study of the $K_0$-group of
 the algebra of a tiling \cite{Con}.
This group plays a role for the \GL\
of \So s which describe discretized particle systems
of solids such as quasicrystals \cite{Be1,BBG} as well as
for topological dynamical systems \cite{HPS,GPS}.

A tiling is a complete covering of $\Real^d$ with distinguished origin $0$ by
\ti s which overlap at most at their boundaries. A \ti\ is closed
subset of $\Real^d$ carrying a \pkt\ and occasionally more decoration.
The \pkt s of the \ti s yield a way to represent a discrete set of allowed
translations, i.e.\ a subset $\tl\2$ of $\Real^d$.
Several \ti s form a \mi\ and we shall be dealing with
\mii es which are equivalence classes of \mi s under translations in $\Real^d$.
Any tiling $\tl$ having for any $r$
only finitely many different \mii es that may be covered by an $r$-ball
defines a compact zero dimensional metric space $\Om$, its hull,
and a (reduced) groupoid-\CA\ $\A_\tl$. The hull is the set of all
tilings which are locally homomorphic to $\tl$ and such that one of their
\pkt s lies on $0$. The groupoid $\Gr\subset\Om\times\Om$ is the groupoid of
the equivalence relation $T\sim T'$ whenever $T'$ is a translate of $T$.
We refer for the details, in particular the topological ones, to \cite{Ke3}
using the notation of that article partly without further explanation.
It is $K_0(\A_\tl)$ together with its natural order structure and scale
which is of interest for the \GL\ and as well for $d$-dimensional topological
dynamical systems (defined by $d$ commuting homeomorphisms on a Cantor
space) which are related to decorations of $\Z^d$.
In the latter case $\Gr$ is isomorphic as a topological groupoid to
a transformation group $\Om\times_\varphi\Z^d$, $\varphi$ denoting the action
of
$\Z^d$ on $\Om$ by homeomorphisms, and the algebra is the
crossed product $C(\Om)\times_\varphi \Z^d$. The Pimsner Voiculescu six term
sequence is the appropriate tool to compute $K$-groups of such
crossed products, and this has been carried out up to $d=3$ \cite{Els2}.
Using the concept of a reduction of a tiling, which is roughly speaking
the same tiling but with a fewer \pkt s,
it could be generalized to tilings (minimal and
of dimension smaller or equal $3$)
which are obtained by the so-called generalized dual method \cite{SoSt2}
from regular grids \cite{Ke3}. In this case a subset $\tl_r\2$ of $\tl\2$
could be singled out by taking those \pkt s which belong to \ti s of a
chosen \mii\ and a continuous action $\varphi$ of $\Z^d$ on
$\Om_r=\overline{\{\tl-x|x\in\tl_r\2\}}$ could be defined so that
$C(\Om_r)\times_\varphi \Z^d$ is stably isomorphic to $\A_\tl$.
This includes tilings obtained from the cut and projection method \cite{DuKa}
hence in particular the Penrose tilings.
To state part of the result
recall that for an \mii\ $M$ with a \pkt\ of one of its tiles $x\in M\2$
\be
\Ch_{M,x}(T) = \left\{
\begin{array}{cl}
1 & \mbox{if $(M,x)$ is at $(T,0)$} \\
0 & \mbox{else}
\end{array}
\right.
\ee
is a continuous function in the topology of $\Om$
where $(M,x)$ is at $(T,0)$ means that $T$ contains a \mi\ of class $M$
such that \pkt $x$ coincides with $0$.
In particular it is a characteristic function on an open and closed set.
Moreover the family of those sets generate the topology of the hull so that
the above functions generate the group $\C$ of continuous functions with
integer values.
Let
$\E$ be the subgroup which is generated by elements of the
form $\Ch_{M,x}-\Ch_{M,x'}$, $x,x'\in M\2$.
\bt
Let $\tl$ be a minimal tiling obtained by the generalized dual method
from a $d$-dimensional regular grid with $d\leq 3$. Then
\be
\begin{array}{lcl}
K_0(\A_\tl)=\C/\E & \mbox{\em for} & d=1 \\
K_0(\A_\tl)=\C/\E\oplus \Z & \mbox{\em for} & d=2\\
K_0(\A_\tl)=\C/\E\oplus H & \mbox{\em for} & d=3.
\end{array}
\ee
\et

{\em Proof:}
The theorem partly summarizes next to the well known result for $d=1$
the content of \cite{Ke3}. It is a consequence of
$C(\Om_r)\times_\varphi \Z^d$ being stably isomorphic to $\A_\tl$ and makes
use of the results of \cite{Els2}.
The only equality which
has still to be shown is $\C/\E=C(\Om_r,\Z)/E_\varphi$ where
$E_\varphi$ is generated by elements of the form $f-f\circ\varphi(m)$
with $m\in \Z^d$ and $f\in C(\Om_r,\Z)$.
It is clear that the inclusion
$C(\Om_r,\Z)\subset \C$ induces an embedding of the quotients so that
only surjectivity has to be shown. From
the minimality of the tiling (which implies that the grid has to be two-sided)
follows that there is
an $R$ such that whenever an \mii\ is
large enough in that it covers an $R$-ball
it has to contain a \ti\ which belongs to the chosen \mii\ of \ti s
that carry a \pkt\ in the reduction of the tiling.
For $M$ large enough let $x_r\in M\2$ be
such a \pkt, then
$\Ch_{M,x}\sim_{\E}\Ch_{M,x_r}$ and $\Ch_{M,x_r}\in C(\Om_r,\Z)$.
Since $\C$ is already generated by functions $\Ch_{M,x}$ with large
enough patterns the result follows.
\eb\bs

The extra summand $\Z$ for $d=2$ is related to
nontrivial vector bundles over the $2$-torus.
In fact, $\A_\tl$ contains
the algebra of continuous complex functions over the $2$-torus and
the generator of the above $\Z$ can be identified with the projection onto a
line bundle over it with chern number $1$.
We do not give the expression for $H$ which is more complicated but known.
It explicitly involves the reduction of the tiling. \bs

We will concentrate here on the computation of $\KT$
for tilings which are invariant
under a locally invertible primitive \sst\ which forces its \saum.\ms

\section{The group $C(\Om,\Z)/\E$ for \sst\ tilings}

A \sst\ of a tiling may be thought of as a rule according to which the
\ti s of the tiling
are to be replaced by \mi s which fit together to yield a new tiling.
It should, up to rescaling, act covariantly with respect to translation,
see \cite{Ke2} for the details.
It may as well be understood as a deflation
(or generalized decomposition \cite{GrSh}) followed by a rescaling of the
\ti s to original size.
A \sst\ tiling is a tiling which is invariant
up to a translation under a \sst.
This expresses a (rather
strong) kind of self-similarity. The \mi\ which replaces a \ti\ or a \mi\ is
called a \fl.

A \sst\ $\rho$ is locally invertible if one can recognize
\fl s by inspection of finite patches (recognizability).
This implies that the functions defined for arbitrary \mii es $M$ and
$x\in\rho^n(M)\2$
\be
\chi^{(n)}_{M,x}(T) = \left\{
\begin{array}{cl}
1 & \mbox{if $(\rho^n(M),x)$ is an $n$-fold \fl\ at $(T,0)$} \\
0 & \mbox{else}
\end{array}
\right.
\ee
are continuous.
That $(\rho^n(M),x)$ is an $n$-fold \fl\ at $(T,0)$ means not only that
$(\rho^n(M),x)$ is at $(T,0)$ but that it occurs  there as an
actual $n$-fold \fl\ of $M$.
A locally invertible \sst\ can be used to construct a
continuous map from the hull $\Om$ onto a path space
$\pf$ over the graph $\Sigma$ which has the \sst\ matrix $\sigma$ as its
 connectivity matrix. This map is surjective if the \sst\ is primitive,
i.e.\ if $\rho^n(a)$ will for any \ti\ $a$ and some $n$ contain all other
\ti s.
Furthermore one obtains an embedding $i:\AF\rightarrow\A_\tl$
of the $AF$-algebra $\AF$
naturally assigned to that path space into the algebra of the tiling.
This embedding induces a homomorphism
$i_*$ of $K$-groups which paves the way to compute $K_0(\A_\tl)$ by
determining $K_0(\AF)$ as well as $\ker i_*$ and $\mbox{coker}\,i_*$.
$K_0(\AF)$ is computed using the functorial
properties of $K_0$ using the fact that $\AF$ is the closure of direct limit
of finite dimensional $C^*$-algebras, see for instance \cite{Mur}.

A \sst\ forces
its \saum\ if there is an $N$ such that whenever $\rho^N(a)$ appears
in $\tl$ as a $N$-fold \fl\ its \saum, i.e.\ the pattern of \ti s which
touch it but do not belong to it, are always the same.
For simplicity we may assume that
$N=1$ as otherwise we could just work with $\rho^N$ as  \sst.
The main reason for the introduction of the
\saum\ forcing condition in \cite{Ke2}
was that it guaranteed the mapping from the hull
$\Om$ onto the path space $\pf$ to be homeomorphic, provided the \sst\ is
primitive.
It follows that the tiling is minimal and
that $\mbox{im}\,i_*=C(\Om,\Z)/\E$.
The generators of $E_\Sigma$ are then given by
$\chi^{(n)}_{M,x}-\chi^{(n)}_{M,x'}$.
In this work we want
to compute $\ker i_*$ which is $\E/E_\Sigma$,
the kernel of the natural projection
$C(\Om,\Z)/E_\Sigma\rightarrow (C(\Om,\Z)/E_\Sigma)/(\E/E_\Sigma)$. \bs


For an \mii\ $M$ we denote by
$\saup{}{M}$ the \mii\ of
$\rho(M)$ together with its \saum, and by $\saup{n}{M}$ the \mii\ of
$\rho^{n-1}(\saup{}{M})$. A $2$-\mi\ is a pattern which consists of two \ti s
which have a common
hypersurface.
We denote $2$ \mi s together with one chosen \pkt\  by $A=(|A|,x)$,
i.e.\ $|A|$ is the corresponding \mii\ and $x\in |A|\2$,
and write $t(A)$ for the \ti\ which carries the \pkt\ $x$.
$[\cdot]$ indicates equivalence classes with respect to $E_\Sigma$.
\bl
Let $\tl$ be invariant under a (primitive)
locally invertible \sst\ which forces its
\saum. Then
$\E/E_\Sigma$ is generated by elements of the form
$[\chi^{(n)}_{M,x}]-[\chi^{(n)}_{M,x'}]$, $x,x'\in \rho^n(M)\2$
where $M$ is a $2$ \mi.
\el
{\em Proof:}
It is clear that a generating set for $\E$ is provided by the set of
elements of the form
$\Ch_{M,x}-\Ch_{M,x'}$ where $M$ is arbitrary and $x,x'\in M\2$
are \pkt s of tiles which have a common hypersurface.
Given such $M,x,x'$ consider the set
\be
I^{(n)}_{M,x,x'}:=\{(A,y)|(M,x)\subset (\saup{n}{|A|},y),
y\in \rho^n(t(A))\2,y\!-\!x\!+\!x'\in \rho^n(|A|)\2\}
\ee
where $A=(A,x'')$ is a $2$ \mi\ with chosen \pkt,
$(M,x)\subset (\saup{n}{|A|},y)$ means that the \mii\ ocurrs inside
$\saup{n}$ such that \pkt\ $x$ coincides with $y$,
 and we keep track
of the position of $\rho^n(t(A))$ and $\rho^n(|A|)$ in $\saup{n}{|A|}$
to whether a \pkt\ of $\saup{n}{|A|}$ lies in $\rho^n(t(A))$ or $\rho^n(|A|)$.
Since  $\rho$ determines its \saum\ we have for big enough $n$
\be
[\Ch_{M,x}]-[\Ch_{M,x'}]=\sum_{(A,y)\in
I^{(n)}_{M,x,x'}}[\chi^{(n)}_{|A|,y}]-[\chi^{(n)}_{|A|,y\!-\!x\!+\!x'}]
\ee
which proves the lemma.\eb\bs

Using the shorter notation
$\chi^{(n)}_i:=[\chi^{(n)}_{a_i,x}]$
with some $x\in \rho^n(a_i)\2$ and $\chi^{(n)}_{A}:=[\chi^{(n)}_{|A|,x}]$
with $x\in \rho^n(t(A))\2$ the group
$C(\Om,\Z)/E_\Sigma$ is generated by
$\{\chi^{(n)}_{i}\}_i$ and
the result of the above lemma can be formulated
by saying that $\E/E_\Sigma$ is generated by elements of the form
$\chi^{(n)}_{A}-\chi^{(n)}_{\bar{A}}$ where $\bar{A}=(A,\bar{x})$ and
$\bar{x}$ is the \pkt\ of the \ti\ which is not $t(A)$.
Recall the definition of the \sst\ matrix
\be
\sigma_{ij} = \hbox{number of $a_j$'s in $\rho(a_i)$}
\ee
which satisfies
\be \label{07022}
\chi^{(n)}_j = \sum_i \chi^{(n+1)}_i \sigma_{ij}.
\ee
In a similar spirit let
\be \label{07021}
K_{i\,A}:=\mbox{number of $A$'s in $F(a_i)$ with $t(A)$
in $\rho(a_i)$}
\ee
for which we keep track of the position of $\rho(a)$ in $F(a)$.
Then
\be
\chi^{(n)}_{A} = \sum_i  \chi^{(n+\6)}_i K_{i\,A}.
\ee
Define
\be
{\cal K}_{i\,A} := K_{i\,A}-K_{i\,\bar{A}}
\ee
and ${\cal K}^{(n)}:\Z^s\rightarrow\KF:
e_{A}\mapsto \sum_i\chi^{(n)}_i {\cal K}_{i\,A}$, where
$s$ is the number of $2$ pattern classes together with a chosen
\pkt. Note that the matrix
$\sigma^n{\cal K}$ plays the role of ${\cal K}$ for the
\sst\ $\rho^{n+1}$, $n\geq 0$.
\bt \label{08021}
Let $\tl$ be a tiling with locally invertible 
\sst\ which forces its \saum.
Then $\E/E_\Sigma$ is generated by $\bigcup_n\mbox{\rm im}\,{\cal K}^{(n)}$.
If moreover the \sst\ matrix $\sigma$ is invertible over $\Z$ so that
$\KF\cong \Z^r$ then
\be \label{03021}
C(\Om,\Z)/\E=\Z^r/<\bigcup_{n\leq 0}\mbox{\rm im\,}\sigma^{n}{\cal K}>
\ee
where ${\cal K}:\Z^s\rightarrow\Z^r:
{\cal K}(e_{A})_i={\cal K}_{i\,A}$.
\et
{\em Proof:} Using (\ref{07022}) and the last lemma
the classes of the generators of $\E/E_\Sigma$ may be expressed as
\be
\chi^{(n)}_{A}-\chi^{(n)}_{\bar{A}} =
\sum_i  \chi^{(m)}_i (\sigma^{m-n-\6}{\cal K})_{i\,A}
\ee
where $m$ is any integer which has,
in case that $\sigma$ is not invertible to be greater or equal to $n+\6$.
Taking $m=n+1$ one obtains
$\E/E_\Sigma=<\bigcup_n\mbox{\rm im}\,{\cal K}^{(n)}>$.
If $\sigma$ is invertible we may take $m=0$ which implies that
$\E/E_\Sigma$ is generated by
$\bigcup_{n\leq 0}\mbox{\rm im\,}\sigma^{n}{\cal K}$.
\eb\bs

Note that due to the subtraction in the
definition of ${\cal K}$ above one only has to count those $2$ \mi s in
$F(a)$ which cross the boundary of $\rho(a)$ and take into account only
one choice for the \pkt\ in $|A|$.

\section{Applications}

Before we apply Theorem~\ref{08021} to Penrose tilings let us
have a look at the one dimensional situation (\sst\ sequences)
and show that (\ref{03021}) fits into the more general results of \cite{For}.
Taking a suitable power of the \sst, which we suppose to be locally invertible
primitive and to force its \saum,
we may as well assume that the first and last
letters of \fl s are stable under it, i.e.\ that
$\rho(\rho(a)_f)_f=\rho(a)_f$ resp.\ $\rho(\rho(a)_l)_l=\rho(a)_l$,
where $w_f$, $w_l$ denote the first resp.\ last letter of the word $w$.
Let us denote by ${\cal F}$ the indices of the first letters and by
${\cal L}$ those of the last.
The \saum\ forcing condition implies that
there is a bijection $\varphi:{\cal L}\rightarrow{\cal F}$.
This means that whenever
$a_j$ is the last (resp.\ first) letter
of $\rho(a_i)$ then the
letter following (resp.\ standing ahead of) it in the
sequence is $a_{\varphi(j)}$ (resp.\ $a_{\varphi^{-1}(j)}$).
To evaluate (\ref{03021}) we may restrict the consideration
to the $2$ patterns
$a_{j}a_{\varphi(j)}$, $j\in{\cal L}$ and choose the \pkt\ to belong to
$a_j$; we abbreviate them as $[j\varphi(j)]$.
This leads to consider
${\cal K}:\Z^{|{\cal L}|}\rightarrow\Z^r$,
$r$ being the total number of letters,
which has entries
\be
{\cal K}_{ij} := {\cal K}_{i\,[j\varphi(j)]} =
\left\{
\begin{array}{rl}
0 & \mbox{if $a_{j}\neq \rho(a_i)_l$ and $a_{\varphi(j)}\neq \rho(a_i)_f$} \\
0 & \mbox{if $a_{j} = \rho(a_i)_l$ and $a_{\varphi(j)} = \rho(a_i)_f$} \\
1 & \mbox{if $a_{j} = \rho(a_i)_l$ and $a_{\varphi(j)}\neq \rho(a_i)_f$} \\
-1 & \mbox{if $a_{j}\neq \rho(a_i)_l$ and $a_{\varphi(j)} = \rho(a_i)_f$}
\end{array}
\right.
\ee
$j\in{\cal L}$. Since $\sigma{\cal K}$ plays the role of ${\cal K}$ but
for $\rho^2$ the above stability
of first and last letters under \sst\ implies
\be
{\cal K}=\sigma{\cal K}.
\ee
Now let us take a piece of the tiling
$\rho(a_{i_1})\rho(a_{i_2})\cdots\rho(a_{i_n})$
and obtain from it a sequence of last
letters $j_1j_2\cdots j_n$, $j_k=\rho(a_{i_k})_l$. If $n$ is large enough
this sequence will by primitivity of the \sst\ contain all $j\in{\cal L}$.
But two consecutive elements $j_{k-1} j_{k}$ of the sequence indicate
that (in vector notation)
${\cal K}_{i_k}=e_{j_{k}}\!-\!e_{j_{k-1}}$ so that the occurrence of
all $j\in{\cal L}$ implies that
$\mbox{im}\,{\cal K}$ is
is generated by $\{e_{j}\!-\!e_{j'}\}_{j,j'\in{\cal L}}$.
Hence in case $\sigma$ is invertible we get
$\mbox{coker}\,{\cal K} = \Z^{r-|{\cal L}|+1}$.

\subsection{$K_0(\A_\tl)$ for Penrose tilings}

There are several well known
variants of tilings which are called Penrose tilings
and which are a priori
to be distinguished as they lead to non-isomorphic groupoids.
The version which is most suitable for our purposes is the one which has
triangles as \ti s, cf.\ Figure~2.
The triangles are decorated (with a little circle) to break the mirror
symmetry.
We may take the centers of these circles to be the \pkt s.
These Penrose tilings by triangles possess an orientational
symmetry of $20$ elements
which is generated by a rotation around $\frac{\pi}{5}$ together with a
mirror reflection at a boundary line of a triangle \cite{Ke2}.
A Penrose tiling by triangles has $40$ \mii es of \ti s.
A short glance at such a tiling shows that always
two triangles of the same congruence
class (under all Euclidian transformations) yield a
$2$ \mi\ which is mirror symmetric and forms a rhombus.
Deleting the corresponding lines yields a tiling made from rhombi.
As a consequence, such a Penrose tiling by rhombi has only $20$ \mii es
of \ti s.
A Penrose tiling by rhombi is actually a reduction of a Penrose tiling
by triangles.
Let
$\tl_{Rh}\2$ consist of all \pkt s which belong to \ti s that
are obtained from those of Figure~2.1 and 2.3 by a rotation.
This clearly defines a reduction and consequently
the algebra of a Penrose tiling by rhombi is stably
isomorphic to the one of a Penrose tiling by triangles and the corresponding
$K_0$-groups are isomorphic as ordered groups.
In fact, like in the proof of Theorem~1 one verifies that
$C(\Om_{Rh},\Z)/E_{\tl_{Rh}}=\C/\E$ where $E_{\tl_{Rh}}=\E\cap C(\Om_{Rh},\Z)$.
We prefer to work with Penrose tilings by triangles since they allow
for a \sst\
which is covariant under the orientational symmetry, and this may be used to
simplify the computation of $\mbox{im}\,{\cal K}$.

Those tilings which have an exact five-fold
symmetry are invariant under the \sst\ $\rho=\check{\rho}^4$ the
deflation corresponding to $\check{\rho}$ being displayed in
Figure~1 for one orientation.

\epsffile[0 0 400 210]{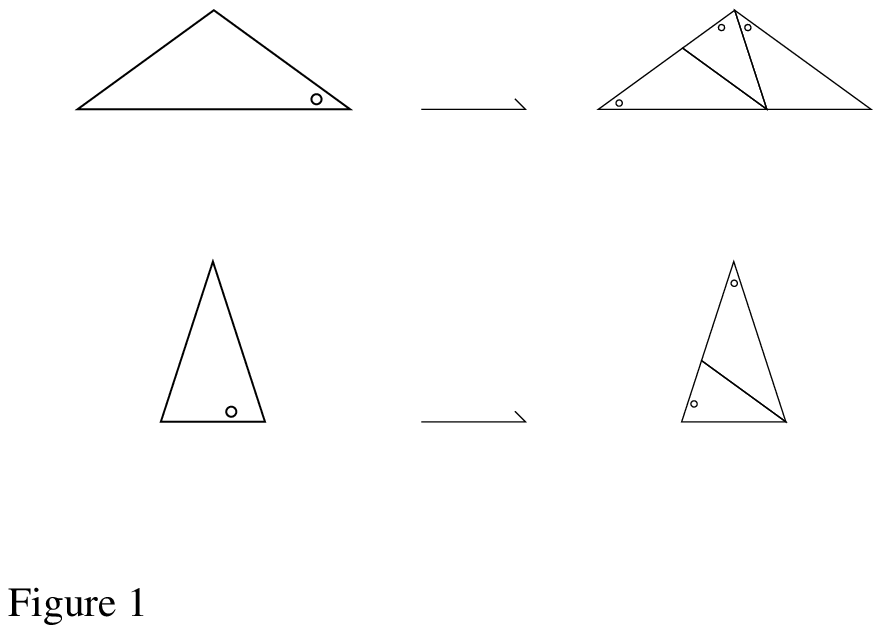}
This \sst\ $\rho$ is primitive, locally invertible,
and forces its
\saum\ with $N=1$ \cite{Ke2}. Its \sst\ matrix is invertible
so that $\C/E_\Sigma=\Z^{40}$.
To simplify the computation of $\E/E_\Sigma$ we make
use of the symmetry properties of the tiling and in particular of the fact
that $2$ patterns crossing the boundaries of $\rho(a)$ are always
mirror symmetric.
Let $\eg$ denote the direction of a boundary line of a tile resp.\ a \fl.
There are
$10$ different ones and we order them  anti-clockwise identifying them with
$\{0,\cdots,9\}$. Saying that a \fl\ has boundary $\eg$ if it has a boundary
with that direction we define
\be
N^\eg_{ij} := \mbox{number of $a_j$'s at boundary $\eg$ of $\rho(a_i)$} .
\ee
In particular $N^\eg_{ij}=0$ in case $a_j$ or $\rho(a_i)$ do not have
boundary $\eg$. Let $a_{\eg(j)}$ be the mirror
image of $a_j$ with respect to the mirror axis $\eg$ and define
\be
{\cal N}^\eg_{ij} = N^\eg_{ij}-N^\eg_{i \eg(j)}
\ee
Then setting
${\cal N}:\Z^{10}\otimes\Z^{40}\rightarrow\Z^{40}:
{\cal N}(e_\alpha\otimes e_j)_i={\cal N}^\eg_{ij}$ we have
\be
{\cal K}_{i\,A} = \left\{
\begin{array}{cl}
{\cal N}^\eg_{ij} & \mbox{if } |A|=a_j a_{\eg(j)}\mbox{ and } t(A)=a_j \\
-{\cal N}^\eg_{ij} & \mbox{if } |A|=a_j a_{\eg(j)}\mbox{ and }t(A)=a_{\eg(j)}\\
0 & \mbox{else}
\end{array}
\right.
\ee
$|A|=a_ja_{\eg(j)}$ indicating that $|A|$ is composed of $a_j$ and
$a_{\eg(j)}$.
Hence
$\mbox{im}\,{\cal K}=\mbox{im}\,{\cal N}$.
${\cal N}^\eg$ is related to ${\cal N}^{0}$ by symmetry, i.e.\
${\cal N}^\eg=R^{-\alpha}{\cal N}^{0} R^\alpha$,
$R$ being the matrix which acts as a rotation around $\frac{\pi}{5}$.
Let us use a basis
$\{\chi^{(0)}_{10 k+\alpha}\}_{0\leq k \leq 3, 0\leq\alpha\leq 9}$
in which $a_{10 k+\alpha}$ corresponds to the \mii\ of the
 triangle in Figure~2.k 
rotated around an angle of $\frac{\alpha\pi}{5}$.

\epsffile[0 0 400 200]{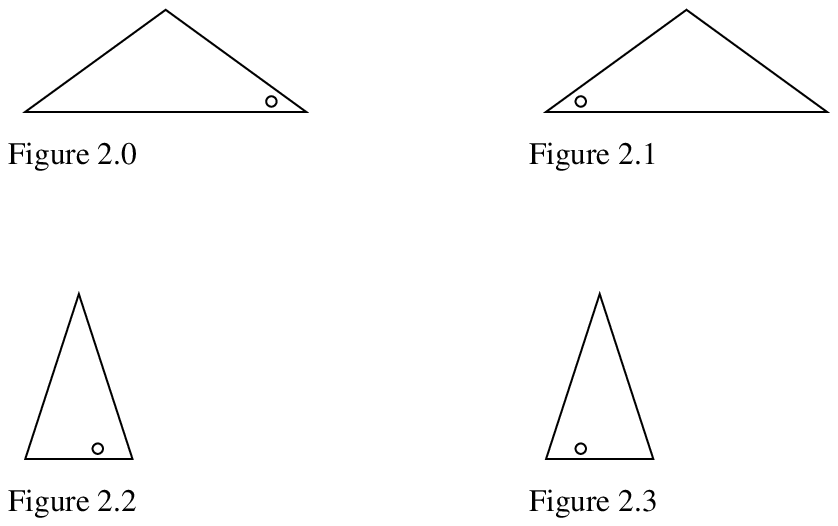}
In terms of the rotation matrix $\omega$
having entries $\omega_{\alpha\beta}=\delta_{\beta-\alpha,1\;mod\;10}$
$R$ and the \sst\ matrix $\sigma=\check{\sigma}^4$ are given by
\begin{eqnarray}
R & = & \left(
\begin{array}{cccc}
\omega & 0 & 0 & 0 \\
0 & \omega & 0 & 0 \\
0 & 0 & \omega & 0 \\
0 & 0 & 0 & \omega
\end{array}
\right) \\
\check{\sigma} & = & \left(
\begin{array}{cccc}
\omega^4 & \omega^0 & 0 & \omega^6 \\
\omega^0 & \omega^6 & \omega^4 & 0 \\
\omega^3 & 0 & \omega^7 & 0 \\
0 & \omega^7 & 0 & \omega^3
\end{array}
\right).
\end{eqnarray}
The matrix $N^{0}$ may be read of from Figure~3 
(and completed by symmetry); it is given below (\ref{19011}).
Moreover ${\cal N}^0=N^0-N^0S$ where $S$ implements the reflection at
$\eg=0$, explicitly, with
$s_{\alpha\beta}=\delta_{\alpha+\beta,5\;mod\;10}$ (counting rows and
columns form $0$ to $9$)
\be
S  =  \left(
\begin{array}{cccc}
0 & s & 0 & 0 \\
s & 0 & 0 & 0 \\
0 & 0 & 0 & s \\
0 & 0 & s & 0
\end{array}
\right).
\ee
It turns out that $\mbox{im}\,{\cal N}$ is generated by
the orbit under $R$ of the four vectors
\be \label{20021} \begin{array}{c}
v_1=(0,\!0,\!0,\!0,\!0,\!0,\!0,\!0,\!0,\!0,\!0,\!0,\!0,\!0,\!0,\!0,\!0,\!0,\!0,
\!0,\!0,\!0,\!0,\!0,\!
   0,\!0,\!1,\!0,\!0,\!0,\!0,\!-1,\!0,\!0,\!0,\!0,
\!0,\!0,\!0,\!0)\\
v_2=(0,\!0,\!0,\!0,\!0,\!0,\!1,\!0,\!0,\!0,\!0,\!-1,\!0,\!0,\!0,
\!0,\!0,\!0,\!0,\!0,\!0,\!0,\!0,\!0,\!
   0,\!0,\!0,\!0,\!0,\!0,\!0,\!0,\!0,\!0,\!0,\!0,\!0,\!0,\!0,\!0)\\
w_1=(0,\!0,\!0,\!0,\!0,\!0,\!0,\!1,\!0,\!0, -1,\!0,\!0,\!0,\!0,
\!0,\!0,\!0,\!0,\!0,\!0,\!0,\!0,\!0,\!
   0,\!0,\!0,\!0,\!1,\!0,\!0,\!0,\!0,\!0,\!0,\!0,\!0,\!0,\!0, -1)\\
w_2=(0,\!0,\!0,\!0,\!0,\!1,\!0,\!0,\!0,\!0,\!0,\!0, -1,\!0,
\!0,\!0,\!0,\!0,\!0,\!0,\!0,\!0,\!0,\!0,\!
   0,\!0,\!0,\!0,\!0, -1,\!0,\!0,\!0,\!0,\!0,\!0,\!0,\!0,\!1,\!0)
\end{array}
\ee
but $8$ of the $40$ vectors thus obtained are linearly dependent.
Moreover $\mbox{im}\,{\cal N}$ is invariant
under $\sigma$. Dividing it out yields no torsion so that we obtain
\be
K_0(\A_\tl)\cong\Z^8\oplus \Z.
\ee
Let us interprete this result. Let $H_1$ be the sublattice of $\Z^{40}$
which is generated by $\{R^\eg v_i\}_{i=1,2;\,\eg=0,1\cdots,9}$.
$(C(\pf,\Z)/E_\Sigma)/H_1=\Z^{20}$ has a basis with natural geometrical
interpretation.
It is formed by the characteristic functions on the \mii es of rhombi which
are obtained from the triangles as described above.

The Penrose tilings by rhombi as they are given by reducing Penrose tilings
by triangles as above do not have symmetric \ti s.
We may delete the circle in the center of which there is no \pkt.
This reduces also
the orientational symmetry to only $10$ elements (the mirror reflection
interchanges
\pkt s from $\tl_{Rh}\2$ with those from $\tl\2\backslash\tl_{Rh}\2$).
But to our knowledge there is no \sst\ for Penrose
tilings by rhombi which is covariant under the orientational symmetry.
But there are at least $10$ non covariant ones;
namely $\rho_\eg=\check{\rho}^4_\eg$
the deflation
corresponding to $\check{\rho}_0$ being given in Figure~4 
and $\check{\rho}_\eg$
being obtained from $\check{\rho}_0$ just by rotation of the whole figure
around $\frac{\eg\pi}{5}$, $\eg=0,\cdots,9$.
That all these
\sst s are primitive locally invertible and force their \saum\ carries over
from $\rho$.
In the basis corresponding to the rhombi ordered as in Figure~4
from above to below
the \sst\ matrix of $\check{\rho}_0$ is given by
\be \label{27021}
\check{\sigma}_0=
\left(\begin{array}{c}
      0\;0\;0\;0\;0\;1\;1\;0\;0\;0\; 0\;0\;0\;0\;0\;0\;0\;0\;0\;0\\
      0\;0\;0\;0\;0\;1\;1\;0\;0\;0\; 0\;0\;0\;0\;0\;0\;0\;0\;0\;0\\
      0\;0\;0\;0\;0\;0\;0\;1\;0\;0\; 0\;1\;0\;1\;0\;0\;0\;0\;0\;0\\
      0\;0\;0\;0\;0\;0\;0\;1\;1\;0\; 0\;0\;1\;0\;0\;0\;0\;0\;0\;0\\
      1\;0\;0\;0\;0\;0\;0\;0\;1\;1\; 0\;0\;0\;0\;0\;0\;0\;0\;0\;0\\
      1\;0\;0\;0\;0\;0\;0\;0\;0\;1\; 0\;0\;0\;0\;1\;0\;1\;0\;0\;0\\
      0\;1\;1\;0\;0\;0\;0\;0\;0\;0\; 0\;0\;0\;0\;0\;1\;0\;1\;0\;0\\
      0\;1\;1\;1\;0\;0\;0\;0\;0\;0\; 0\;0\;0\;0\;0\;0\;0\;0\;0\;0\\
      0\;0\;0\;1\;1\;0\;0\;0\;0\;0\; 0\;0\;0\;0\;0\;0\;0\;0\;0\;1\\
      0\;0\;0\;0\;1\;0\;0\;0\;0\;0\; 1\;0\;0\;0\;0\;0\;0\;0\;1\;0\\
      0\;0\;0\;0\;0\;0\;0\;0\;0\;0\; 0\;0\;0\;1\;0\;0\;0\;0\;0\;0\\
      0\;0\;0\;0\;0\;0\;0\;0\;0\;0\; 0\;0\;0\;0\;0\;0\;0\;0\;1\;0\\
      0\;0\;0\;0\;0\;1\;0\;0\;0\;0\; 0\;0\;0\;0\;0\;0\;0\;0\;0\;0\\
      1\;0\;0\;0\;0\;0\;0\;0\;0\;0\; 0\;0\;0\;0\;0\;0\;1\;0\;0\;0\\
      0\;0\;0\;0\;0\;0\;0\;1\;0\;0\; 0\;1\;0\;0\;0\;0\;0\;1\;0\;0\\
      0\;0\;1\;0\;0\;0\;0\;0\;1\;0\; 0\;0\;1\;0\;0\;0\;0\;0\;0\;0\\
      0\;0\;0\;1\;0\;0\;0\;0\;0\;1\; 0\;0\;0\;0\;0\;0\;0\;0\;0\;1\\
      0\;0\;0\;0\;1\;0\;0\;0\;0\;0\; 1\;0\;0\;0\;1\;0\;0\;0\;0\;0\\
      0\;1\;0\;0\;0\;0\;0\;0\;0\;0\; 0\;0\;0\;0\;0\;1\;0\;0\;0\;0\\
      0\;0\;0\;0\;0\;0\;1\;0\;0\;0\; 0\;0\;0\;0\;0\;0\;0\;0\;0\;0
\end{array}\right).
\ee
It turns out that
\be
\ker\check{\sigma}_\eg = <\{R^\eg w_i\}_{i=1,2}>.
\ee
Since the restriction to its image is an automorphism and
$\ker\check{\sigma}_0=\ker\check{\sigma}^2_0$ we have
\be
C({\cal P}_{\Sigma_\eg},\Z)/E_{\Sigma_\eg}
\cong \Z^{20}/
<\{R^\eg w_i\}_{i=1,2}>
\ee
$\Sigma_\eg$ denoting the graph having $\check{\sigma}_\eg$ as connectivity
matrix. Let $j:\Z^{20}\rightarrow C(\Om_{Rh},\Z)/E_{\tl_{Rh}}$ be given by
assigning the characteristic functions onto the \mii es of rhombi
to the standard basis. It is surjective since $H_1\subset\mbox{im}\,{\cal N}$.
Any of the \sst s $\rho_\eg$ leads to a homeomorphism between
$\Om_{Rh}$ and the path space ${\cal P}_{\Sigma_\eg}$ defined by $\Sigma_\eg$
and moreover to an embedding $i_\eg:\A_{\Sigma_\eg}\rightarrow \A_\tl$
of the $AF$-algebra naturally assigned to ${\cal P}_{\Sigma_\eg}$
into the algebra of $\tl$.
Let $\pi_\eg$ be the natural projection onto $\Z^{20}/\ker\sigma_\eg$
then $j=(i_\eg)_*\circ \pi_\eg $ for all $\eg$ and therefore
\be
\ker j \subset <\{R^\eg w_i\}_{i=1,2;\,\eg=1,\cdots,9}>\cong \Z^{12}.
\ee
This shows independently that
$\C/\E\subset\Z^{8}$ but not the opposite inclusion.
A computation of $\mbox{im}\,{\cal K}$ for e.g.\ $\rho_0$
would have been more complicated due to the lack of symmetry.\bs

It remains to compute the order structure.
The above result for the $K_0$-group raises
the question  whether
there is a tiling which is mutually locally derivable from a
Penrose tiling but has only $8$ \mii es of \ti s.\bs

 \newpage
\be \label{19011}
N^{0} =
\left(\begin{array}{c}
3\;0\;0\;0\;0\;0\;0\;0\;0\;0\;1\;0\;0\;0\;1\;0\;2\;0\;0\;0\;0\;0\;0\;0\;0\;0\;0
\;1\;0\;0\;1\;0\;0\;0\;0\;0\;
   0\;0\;0\;0\\ 0\;1\;0\;0\;0\;0\;0\;0\;0\;1\;0\;0\;0\;0\;0\;2\;0\;0\;0\;0\;0
\;0\;0\;0\;0\;1\;0
\;0\;0\;0\;
   0\;0\;0\;0\;0\;0\;0\;0\;1\;0\\
0\;0\;0\;0\;0\;0\;0\;0\;0\;0\;0\;0\;0\;0\;0\;0\;0\;0\;0\;0\;0\;0\;0\;0\;
   0\;0\;0\;0\;0\;0\;0\;0\;0\;0\;0\;0\;0\;0\;0\;0\\
   0\;0\;0\;0\;0\;0\;0\;0\;0\;0\;0\;0\;0\;0\;0\;0\;0\;0\;0\;0\;0\;0\;0\;0\;0
\;0\;0
\;0\;0\;0\;0\;0\;0\;0\;0\;0\;
   0\;0\;0\;0\\ 1\;0\;0\;0\;1\;0\;1\;0\;0\;0\;1\;0\;0\;0\;0\;0\;1\;0\;0
\;0\;1\;0\;0\;0\;0
\;0\;0\;0\;0\;0\;
   0\;0\;0\;0\;0\;0\;0\;0\;0\;0\\
0\;0\;0\;0\;0\;3\;0\;0\;0\;0\;0\;2\;0\;0\;0\;1\;0\;0\;0\;1\;0\;0\;1\;0\;
   0\;0\;0\;0\;0\;0\;0\;0\;0\;0\;0\;1\;0\;0\;0\;0\\
   0\;0\;0\;0\;1\;0\;1\;0\;0\;0\;2\;0\;0\;0\;0\;0\;0\;0\;0\;0\;1\;0\;0\;0\;0
\;0\;0\;0\;0\;0\;0\;0\;0\;1\;0\;0\;
   0\;0\;0\;0\\ 0\;0\;0\;0\;0\;0\;0\;0\;0\;0\;0\;0\;0\;0\;0\;0\;0\;0\;0\;0
\;0\;0\;0\;0\;0\;0
\;0\;0\;0\;0\;
   0\;0\;0\;0\;0\;0\;0\;0\;0\;0\\
0\;0\;0\;0\;0\;0\;0\;0\;0\;0\;0\;0\;0\;0\;0\;0\;0\;0\;0\;0\;0\;0\;0\;0\;
   0\;0\;0\;0\;0\;0\;0\;0\;0\;0\;0\;0\;0\;0\;0\;0\\
   0\;1\;0\;0\;0\;1\;0\;0\;0\;1\;0\;1\;0\;0\;0\;1\;0\;0\;0\;0\;0\;0\;0\;0\;0
\;1\;0\;0\;0\;0\;0\;0\;0\;0\;0\;0\;
   0\;0\;0\;0\\ 1\;0\;0\;0\;2\;0\;1\;0\;0\;0\;3\;0\;0\;0\;0\;0\;0\;0\;0\;0
\;1\;0\;0\;0\;0\;0
\;0\;0\;0\;0\;
   0\;0\;0\;1\;0\;0\;0\;0\;0\;0\\
0\;0\;0\;0\;0\;1\;0\;0\;0\;1\;0\;1\;0\;0\;0\;1\;0\;0\;0\;1\;0\;0\;0\;0\;
   0\;0\;0\;0\;0\;0\;0\;0\;0\;0\;0\;1\;0\;0\;0\;0\\
   0\;0\;0\;0\;0\;0\;0\;0\;0\;0\;0\;0\;0\;0\;0\;0\;0\;0\;0\;0\;0\;0\;0\;0\;0\;0
\;0\;0\;0\;0\;0\;0\;0\;0\;0\;0\;
   0\;0\;0\;0\\ 0\;0\;0\;0\;0\;0\;0\;0\;0\;0\;0\;0\;0\;0\;0\;0\;0\;0\;0\;0
\;0\;0\;0\;0\;0
\;0\;0\;0\;0\;0\;
   0\;0\;0\;0\;0\;0\;0\;0\;0\;0\\
2\;0\;0\;0\;0\;0\;0\;0\;0\;0\;0\;0\;0\;0\;1\;0\;1\;0\;0\;0\;0\;0\;0\;0\;
   0\;0\;0\;1\;0\;0\;1\;0\;0\;0\;0\;0\;0\;0\;0\;0\\
   0\;1\;0\;0\;0\;1\;0\;0\;0\;2\;0\;0\;0\;0\;0\;3\;0\;0\;0\;0\;0\;0\;0\;0\;0
\;1\;0\;0\;0\;0\;0\;0\;0\;0\;0\;0\;
   0\;0\;1\;0\\ 1\;0\;0\;0\;1\;0\;0\;0\;0\;0\;1\;0\;0\;0\;1\;0\;1\;0\;0\;0
\;0\;0\;0\;0\;0
\;0\;0\;0\;0\;0\;
   1\;0\;0\;0\;0\;0\;0\;0\;0\;0\\
0\;0\;0\;0\;0\;0\;0\;0\;0\;0\;0\;0\;0\;0\;0\;0\;0\;0\;0\;0\;0\;0\;0\;0\;
   0\;0\;0\;0\;0\;0\;0\;0\;0\;0\;0\;0\;0\;0\;0\;0\\
   0\;0\;0\;0\;0\;0\;0\;0\;0\;0\;0\;0\;0\;0\;0\;0\;0\;0\;0\;0\;0\;0\;0\;0\;0
\;0\;0\;0\;0\;0\;0\;0\;0\;0\;0\;0\;
   0\;0\;0\;0\\ 0\;0\;0\;0\;0\;2\;0\;0\;0\;0\;0\;1\;0\;0\;0\;0\;0\;0\;0\;1
\;0\;0\;1\;0\;0\;0
\;0\;0\;0\;0\;
   0\;0\;0\;0\;0\;1\;0\;0\;0\;0\\
1\;0\;0\;0\;0\;0\;0\;0\;0\;0\;0\;0\;0\;0\;1\;0\;1\;0\;0\;0\;0\;0\;0\;0\;
   0\;0\;0\;0\;0\;0\;1\;0\;0\;0\;0\;0\;0\;0\;0\;0\\
   0\;0\;0\;0\;0\;0\;0\;0\;0\;0\;0\;0\;0\;0\;0\;0\;0\;0\;0\;0\;0\;0\;0\;0\;0\;0
\;0\;0\;0\;0\;0\;0\;0\;0\;0\;0\;
   0\;0\;0\;0\\ 0\;1\;0\;0\;0\;0\;0\;0\;0\;1\;0\;0\;0\;0\;0\;2\;0\;0\;0\;0
\;0\;0\;0\;0\;0\;1
\;0\;0\;0\;0\;
   0\;0\;0\;0\;0\;0\;0\;0\;1\;0\\
1\;0\;0\;0\;1\;0\;0\;0\;0\;0\;1\;0\;0\;0\;1\;0\;1\;0\;0\;0\;0\;0\;0\;0\;
   0\;0\;0\;0\;0\;0\;1\;0\;0\;0\;0\;0\;0\;0\;0\;0\\
   0\;0\;0\;0\;0\;0\;0\;0\;0\;0\;0\;0\;0\;0\;0\;0\;0\;0\;0\;0\;0\;0\;0\;0\;0\;0
\;0\;0\;0\;0\;0\;0\;0\;0\;0\;0\;
   0\;0\;0\;0\\ 0\;0\;0\;0\;0\;1\;0\;0\;0\;0\;0\;1\;0\;0\;0\;0\;0\;0\;0\;1\;0
\;0\;0\;0\;0\;0\;0
\;0\;0\;0\;
   0\;0\;0\;0\;0\;1\;0\;0\;0\;0\\
0\;0\;0\;0\;0\;0\;0\;0\;0\;0\;0\;0\;0\;0\;0\;0\;0\;0\;0\;0\;0\;0\;0\;0\;
   0\;0\;0\;0\;0\;0\;0\;0\;0\;0\;0\;0\;0\;0\;0\;0\\
   0\;0\;0\;0\;1\;0\;1\;0\;0\;0\;2\;0\;0\;0\;0\;0\;0\;0\;0\;0\;1\;0\;0\;0\;0
\;0\;0
\;0\;0\;0\;0\;0\;0\;1\;0\;0\;
   0\;0\;0\;0\\ 0\;0\;0\;0\;0\;1\;0\;0\;0\;1\;0\;1\;0\;0\;0\;1\;0\;0\;0\;1
\;0\;0\;0\;0\;0\;0
\;0\;0\;0\;0\;
   0\;0\;0\;0\;0\;1\;0\;0\;0\;0\\
0\;0\;0\;0\;0\;0\;0\;0\;0\;0\;0\;0\;0\;0\;0\;0\;0\;0\;0\;0\;0\;0\;0\;0\;
   0\;0\;0\;0\;0\;0\;0\;0\;0\;0\;0\;0\;0\;0\;0\;0\\
   0\;0\;0\;0\;1\;0\;1\;0\;0\;0\;1\;0\;0\;0\;0\;0\;0\;0\;0\;0\;1\;0\;0\;0\;0\;0
\;0\;0\;0\;0\;0\;0\;0\;0\;0\;0\;
   0\;0\;0\;0\\ 0\;0\;0\;0\;0\;0\;0\;0\;0\;0\;0\;0\;0\;0\;0\;0\;0\;0\;0\;0\;0
\;0\;0\;0\;0\;0\;0
\;0\;0\;0\;
   0\;0\;0\;0\;0\;0\;0\;0\;0\;0\\
0\;1\;0\;0\;0\;1\;0\;0\;0\;1\;0\;1\;0\;0\;0\;1\;0\;0\;0\;0\;0\;0\;0\;0\;
   0\;1\;0\;0\;0\;0\;0\;0\;0\;0\;0\;0\;0\;0\;0\;0\\
   2\;0\;0\;0\;0\;0\;0\;0\;0\;0\;0\;0\;0\;0\;1\;0\;1\;0\;0\;0\;0\;0\;0\;0\;0
\;0\;0
\;1\;0\;0\;1\;0\;0\;0\;0\;0\;
   0\;0\;0\;0\\ 0\;0\;0\;0\;0\;0\;0\;0\;0\;0\;0\;0\;0\;0\;0\;0\;0\;0\;0\;0
\;0\;0\;0\;0\;0\;0
\;0\;0\;0\;0\;
   0\;0\;0\;0\;0\;0\;0\;0\;0\;0\\
0\;1\;0\;0\;0\;0\;0\;0\;0\;1\;0\;0\;0\;0\;0\;1\;0\;0\;0\;0\;0\;0\;0\;0\;
   0\;1\;0\;0\;0\;0\;0\;0\;0\;0\;0\;0\;0\;0\;0\;0\\
   0\;0\;0\;0\;0\;0\;0\;0\;0\;0\;0\;0\;0\;0\;0\;0\;0\;0\;0\;0\;0\;0\;0\;0\;0\;0
\;0\;0\;0\;0\;0\;0\;0\;0\;0\;0\;
   0\;0\;0\;0\\ 1\;0\;0\;0\;1\;0\;1\;0\;0\;0\;1\;0\;0\;0\;0\;0\;1\;0\;0\;0
\;1\;0\;0\;0\;0\;0
\;0\;0\;0\;0\;
   0\;0\;0\;0\;0\;0\;0\;0\;0\;0\\
0\;0\;0\;0\;0\;2\;0\;0\;0\;0\;0\;1\;0\;0\;0\;0\;0\;0\;0\;1\;0\;0\;1\;0\;
   0\;0\;0\;0\;0\;0\;0\;0\;0\;0\;0\;1\;0\;0\;0\;0\\
   0\;0\;0\;0\;0\;0\;0\;0\;0\;0\;0\;0\;0\;0\;0\;0\;0\;0\;0\;0\;0\;0\;0\;0\;0\;0
\;0\;0\;0\;0\;0\;0\;0\;0\;0\;0\;
   0\;0\;0\;0
\end{array}\right)
\ee
\newpage
\epsffile[0 0 465 615]{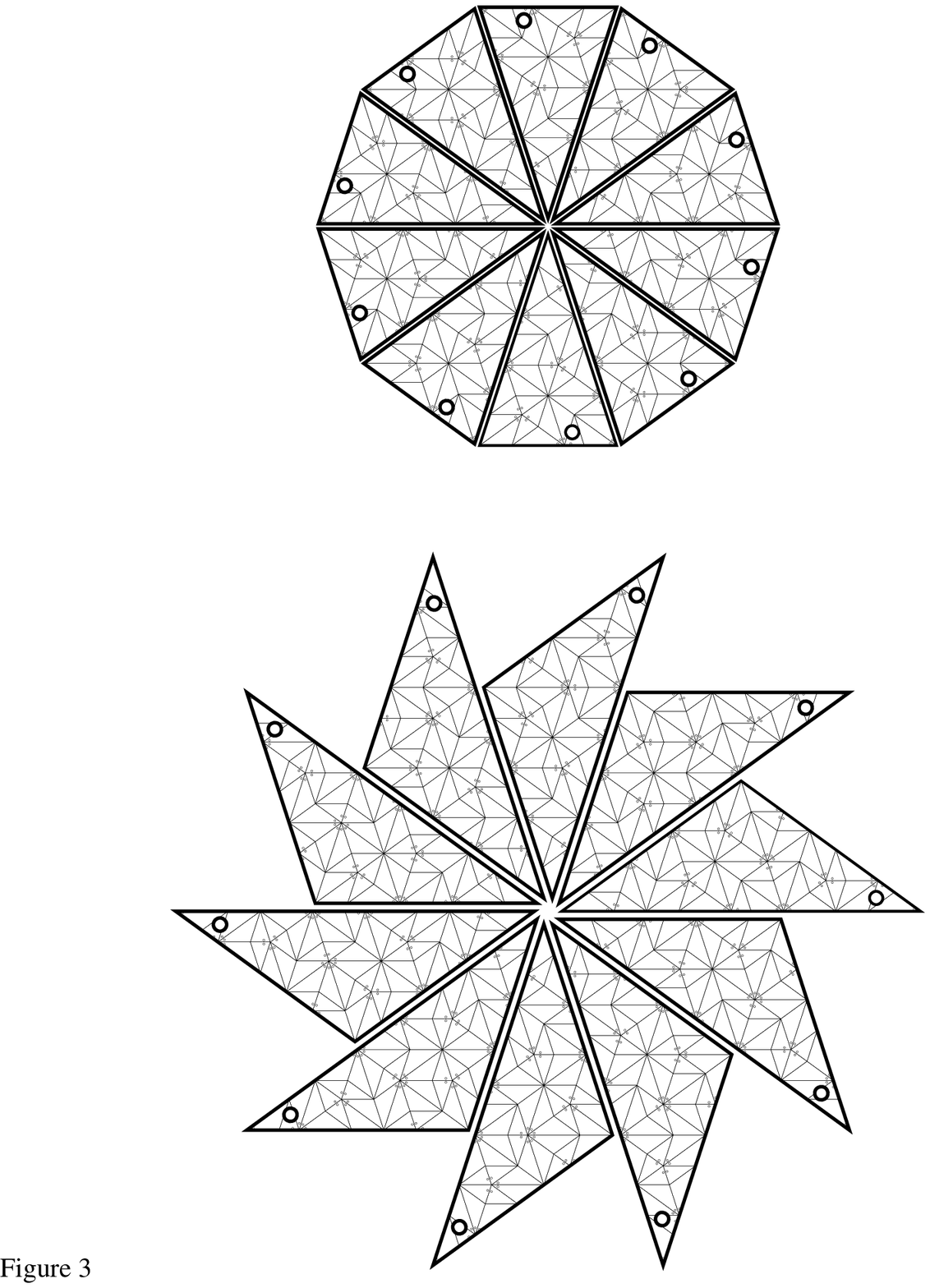}
Substitutes of $\rho=\check{\rho}^4$ of the triangles of Figure~2.0 and
2.1 in all directions.
\newpage
\epsffile[0 0 465 645]{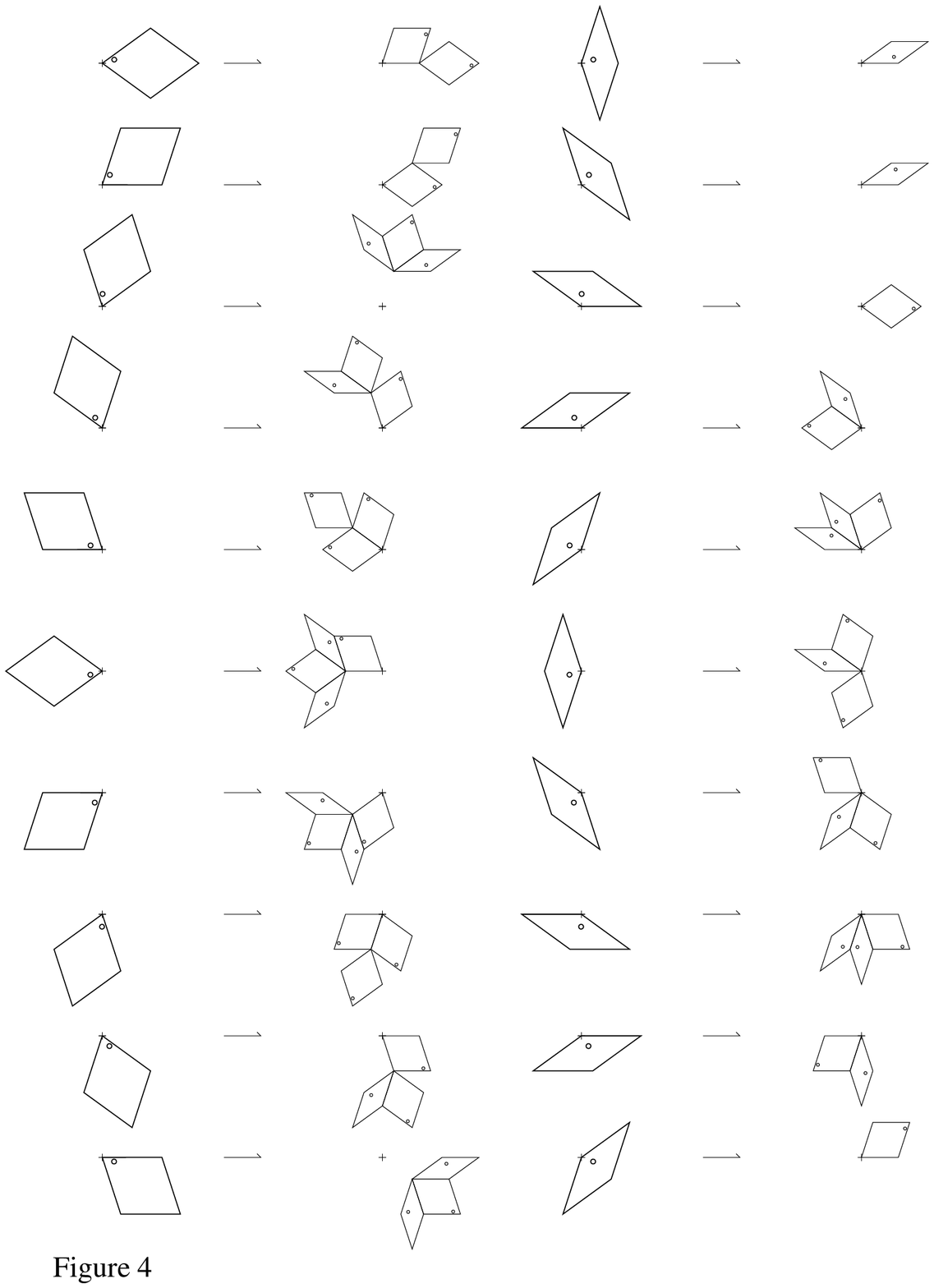}
\newpage

{\bf Acknowledgement.}
I thank Alan Forrest for very fruitful conversations.

\end{document}